\documentclass[prl,amsmath,amssymb, twocolumn, showpacs]{revtex4}

\usepackage{dcolumn} 

\usepackage[dvips]{graphicx}

\usepackage{epsfig}

\begin{document}

\title{Supersymmetry in quantum mechanics: An extended view}

\author{A.\ R.\  P. Rau$^{*}$}
\affiliation{Department of Physics and Astronomy, Louisiana State University,
Baton Rouge, Louisiana 70803-4001}


\begin{abstract}

The concept of supersymmetry in a quantum mechanical system is extended, permitting the recognition of many more supersymmetric systems, including very familiar ones such as the free particle. Its spectrum is shown to be supersymmetric, with space-time symmetries used for the explicit construction. No fermionic or Grassmann variables need to be invoked. Our construction extends supersymmetry to continuous spectra. Most notably, while the free particle in one dimension has generally been regarded as having a doubly degenerate continuum throughout, the construction clarifies that there is a single zero energy state at the base of the spectrum. 
\end{abstract}


\maketitle

Supersymmetry (SUSY) in field theories of elementary particles results in a fermion partner to every boson, and vice versa.  Both partners have the same mass (synonymously, energy). The resulting pattern has one non-degenerate, zero-energy, bosonic state (the vacuum), while every other state of the spectrum is degenerate in pairs (a boson and a fermion). This is termed a SUSY spectrum \cite{ref1}. There is as yet no experimental evidence of such SUSY particles although certain instances of the spectra of neighboring nuclei have been claimed as examples of approximate SUSY in nature \cite{ref2}. Witten \cite{ref3} extended the SUSY concept to ``zero-dimensional field theories", that is, to ordinary quantum mechanics. The many examples studied \cite{ref4} have provided concrete physical systems and much interesting insight into SUSY. In this Letter, the concept is extended still further. Examples, including some very simple and familiar systems and systems with purely continuous spectra, show the existence of many more SUSY systems. Remarkably, it seems to have gone unrecognized that the free particle itself has SUSY. The supersymmetry construction provides further insight into the nature of this simplest of all physical systems. Instead of invoking fermionic spin variables from the start, our construction uses space-time symmetries such as parity and time reversal.  

Fig.\ 1 gives a schematic of a SUSY spectrum. Among the distinguishing characteristics are \cite{ref4} : 1) a zero energy non-degenerate ground state, 2) pairs of degenerate excited states, 3) the existence of operators $Q$ and $Q^{\dagger}$ that carry degenerate pairs into one another as shown while both operators yield zero upon acting on the ground state, 4) the operators satisfy

\begin{equation}
[H,Q]=[H,Q^{\dagger}]=0, H=\{Q,Q^{\dagger}\}/2,
\label{eqn1}
\end{equation}
5) the operators have fermionic or ``Grassmann" character in that

\begin{equation}
\{Q,Q\}=\{Q^{\dagger},Q^{\dagger}\}=0,
\label{eqn2}
\end{equation}
and 6) as seen in the above equations, the operators involved form a closed algebra with mixed commutators and anti-commutators. SUSY quantum mechanics also identifies pairs of potentials, $V_{\pm}$, which separately support the two ladders of states in Fig.\ 1. These potentials can be expressed in terms of the ground state wave function and its first derivative \cite{ref4}.

\begin{figure}
\hspace{-0.6in}
\vspace{-2.2in}
\scalebox{1.8}{\includegraphics[width=2in]{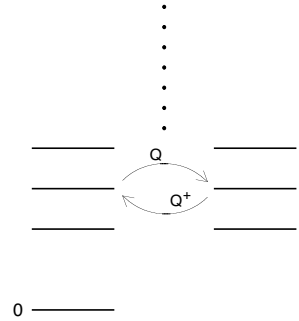}}
\vspace{.7in}
\caption{A schematic SUSY spectrum. Note a non-degenerate ground state of zero energy and an excited state spectrum of degenerate pairs with operators that transform one into the other.}
\end{figure}

Turning now to our extension, consider a free particle of mass $m$ in one dimension, easily the simplest non-relativistic quantum mechanical system and one encountered in our very first acquaintance with the subject. With the wave number $k$ defined in the usual way as $E=(\hbar k)^2/2m$, the spectrum of the Hamiltonian, $H= \frac{1}{2} p^2 =-\frac{1}{2} d^2/dx^2$, is shown in Fig.\ 2. (We will set $\hbar =m=1$.) The purely continuous spectrum has two alternative descriptions. In terms of standing waves, even and odd parity eigenstates are as shown. The corresponding wave functions are $\cos kx$ and $\sin kx$, respectively. Apart from an irrelevant numerical factor, these functions are ``normalized per unit momentum", that is, to a Dirac delta function in wave number (see Sec.\ 21 of \cite{ref5}). There is a general view that the free particle has a doubly-degenerate continuous spectrum but, remarkably, it seems to have escaped notice that this is only true for all $k >0$, the $k=0$ occurring only in the even parity sector. The odd parity state vanishes in this limit. In Fig.\ 2, the bottom of the odd parity continuum is shown dashed to denote the absence of the $k=0$ value. Thus, a free particle has a SUSY spectrum.

\begin{figure}
\vspace{-1.8in}
\hspace{-.65in}
\scalebox{2.0}{\includegraphics[width=2in]{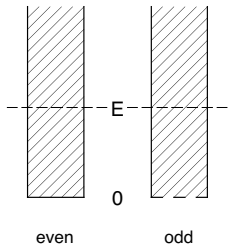}}
\vspace{-1.9in}
\caption{Spectrum of a free particle, showing two continua of even and odd parity. Dashed line at the base of the latter indicates the absence of the zero-energy state.}
\end{figure}

The zero-energy differential equation has two independent solutions, a constant and $x$. But our claim is that only the former corresponds to a physical state, the other solution not being finite at infinity \cite{ref6}. All the other trigonometric functions for non-zero $k$ and the even parity function for $k=0$ are finite everywhere even if normalizable only to a delta function. A supporting argument for a unique state at the base of the continuum is provided by looking at
the alternative description of the free particle in terms of traveling waves, or momentum eigenstates. Wave functions $\exp (\pm ikx)$ denote waves traveling from left to right and vice versa. Again, while such pairs exist for non-zero $k$, the $k=0$ limit that describes no propagation at all is but a single state. So, in either of the two standard descriptions, the free particle spectrum should be viewed as one of SUSY. Indeed, the next paragraph views the free particle as a limiting case of  ``quantization in a box", together with its SUSY construction, providing strong supporting evidence. 

A particle of mass $m$ in a one-dimensional box $(-\frac{1}{2}L, \frac{1}{2}L)$ has eigenvalues $E_n=n^2 (\pi^2 \hbar^2/2mL^2), n=1,2,\ldots$, and its ground state wave function is $\cos(\pi x/L)$. The higher excited states are in sequence, $\sin (2\pi x/L), \cos (3\pi x/L), \ldots$, and alternate in parity. Its SUSY partner, as is well-known \cite{ref7}, has potential $V_{-} = 2(\hbar ^2/2m)(\pi /L)^2 \sec ^2 (\pi x/L)$ with the same eigenvalues except that $n=1$ is missing. The lowest state's wave function is $\cos ^2 (\pi x/L)$, has even parity, while the next excited state of odd parity and energy $E_3$ has wave function $\cos ^2 (\pi x/L) \sin (\pi x/L)$. Note how this entirely discrete SUSY spectrum for any finite $L$ evolves as $L \rightarrow \infty $. Both potentials go to zero, giving the free particle Hamiltonian, and the eigenvalues condense into continuous spectra of even and odd parity pairs but there is always one extra state in $V_{+}$. In this limit, with the spectra of the two potentials superposed and an orthonormal set formed, the free particle spectrum emerges with cosine and sine pairs in $n\pi x/L \rightarrow kx$ for all $n \geq 2$. But the ground $n=1\rightarrow k=0$ state stands by itself.

The free particle's SUSY nature may have escaped notice in part, perhaps, because SUSY in quantum mechanics has generally been discussed for bound states. Also, a normalized wave function of the ground state is often used for constructing the partner potentials $V_{\pm}$. But with continuum normalization to a Dirac delta function, the spectrum of a free particle is undeniably supersymmetric as we now proceed to show by completing the construction of the operators in Eq.~(\ref{eqn1}).  

With the operators of momentum $p$ and parity $P$, we define

\begin{equation}
Q = pP/\sqrt{2m}, Q^{\dagger} = Pp/\sqrt{2m} =-Q,
\label{eqn3}
\end{equation}
which clearly satisfy Eq.~(\ref{eqn1}). They also satisfy criterion 3 in the list above, carrying sines and cosines into one another to within multiplicative constants and giving zero when operating on the constant ground state wave function. Only criterion 5 is not satisfied (more on this below), no fermionic element having been introduced. The squares of the operators do not vanish but yield the Hamiltonian to within a sign. (This sign and the anti-Hermitian nature of $Q$ can be removed by multiplying with an $i$ in Eq.~(\ref{eqn3}) above. This is a matter of taste.) The operators again close under mixed commutators and anti-commutators, thus satisfying criterion 6.

It is interesting that the two alternative descriptions of the free particle are generally associated with the existence of two alternative operators, $p$ and $P$, both of which commute with $H$ while not commuting with one another. Clearly, their product also commutes with the Hamiltonian and this is precisely what Eq.~(\ref{eqn3}) builds on in constructing SUSY. This same procedure will be used below for the rotational counterpart and may afford further generalization.

While the above construction in which criterion 5 is relaxed has some merit, pointing to an extended view of SUSY constructions, even that criterion can be retained by defining an alternative supercharge to the above Q. This is constructed by superposing the momentum operator p with the operator pP:

\begin{equation}
q=(p+pP)/\sqrt{4m}, q^{\dagger}=(p-pP)/\sqrt{4m},
\label{eqn4}
\end{equation}
such that now $q^2=0$ while Eq.~(\ref{eqn1}) is still satisfied. Unlike $Q$, this supercharge $q$ also has the fermionic element of criterion 5 and satisfies Eq.~(\ref{eqn2}) . Its actions on the wave functions are given by

\begin{eqnarray}
q \cos kx & = & i(\hbar k/\sqrt{m}) \sin kx, \, q \sin kx =0 \nonumber \\
q^{\dagger} \sin kx & = & -i(\hbar k/\sqrt{m}) \cos kx, \, q^{\dagger} \cos kx =0.
\label{eqn5}
\end{eqnarray}
Note that both $q$ and $q^{\dagger}$ give zero when acting on the ground state. Although the two constructions through $Q$ or $q$ are equivalent, we find some merit in the former as simpler and more straightforward, with just one operator $pP$ rather than two, $p$ and $pP$. But, such an equivalence in general of $N=1$ and $N=2$ SUSY has been noted \cite{ref8}.

The idea of using as a ``grading" operator an operator such as parity whose square is unity, in contrast to a Grassmann variable or fermionic object whose square vanishes, has occurred before \cite{ref9}. In particular, Plyushchay and co-workers have used reflection operators in a series of papers \cite{ref10} for bosonized SUSY and parabosonic systems. However, in many cases, the reflection operators have occurred in conjunction with Pauli spinors and even the ``minimal" example in the second of these papers uses a superpotential which is an odd function under parity. Our construction in this Letter, using only parity, and with the superpotential and $V_{\pm}$ all equal to zero as pertaining to a free particle, is perhaps the simplest illustration of grading. Our treatment of a continuous spectrum also departs from all other discussions so far of SUSY in quantum mechanics which have only considered bound states, perhaps reflecting the bias of its origins in SUSY field theories.   

Consider next a zero-range, delta function well in one dimension, with $H=p^2/2 -\lambda \delta(x)$. As is well-known, such an attractive delta-well has one and only one bound state with energy $E=-\lambda^2/2$ and normalized wave function $\sqrt{\lambda} \exp(-\lambda |x|)$, and these are often used as a model for loosely bound systems such as the deuteron in nuclear physics or a negative ion in atomic physics.  The entire spectrum is as shown in Fig.\ 3, the odd-parity continuum remaining unchanged (from the free particle's) as $\sin kx$ since these states never ``see" the zero-range potential whereas the even parity states are now given by \cite{ref11}

\begin{equation}
\frac{k \cos kx-\lambda \sin k|x|}{\sqrt{k^2+\lambda^2}}.
\label{eqn6}
\end{equation}
The change from $\cos kx$ is necessary to satisfy the Schr\"{o}dinger equation with the delta-well and to be orthogonal to the ground state wave function. In the limit $\lambda \rightarrow 0$, Eq.~(\ref{eqn6}) reduces, of course, to the free particle $\cos kx$. But now, as emphasized by the dashed lines at the bottom of both continua in Fig.\ 3, the limit $k \rightarrow 0$ makes Eq.~(\ref{eqn6}) vanish. The $k=0$ state is now absent also in the even parity sector. Once again, there is a SUSY spectrum, with one even parity ground state that is bound and all excited states being continuum states in pairs. The zero of the energy scale can be reset at the bound state energy. Comparison with the previous example shows also how one state from the even parity continuum is ``peeled" off to become the bound state upon ``turning on" $\lambda$. This provides yet another perspective on the free particle and delta-well spectrums, both of which exhibit SUSY.

\begin{figure}
\vspace{-1.5in}
\hspace{-.65in}
\scalebox{1.8}{\includegraphics[width=2in]{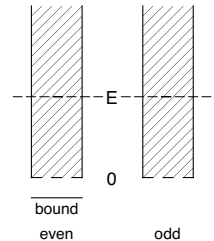}}
\vspace{-1.5in}
\caption{Spectrum of a one-dimensional delta function potential well. Contrast with Fig.\ 2. While the odd parity states are unchanged, there is now a bound state of even parity and the lowest state $k=0$ of the continuum is absent.}
\end{figure}

A related problem that has been discussed earlier \cite{ref13} (see also Fig.\ 8 of Ref. [12]) is the one-dimensional ``cut-off" Coulomb potential that applies to a hydrogen atom in an ultrastrong magnetic field of the sort found on pulsars \cite{ref14}. There is a SUSY spectrum with one deeply bound ground state, all others being even and odd parity pairs at the usual Bohr energies.

The rotational counterpart of the free particle in translation is, of course, the free rotor in a plane with Hamiltonian $L_z^2/2I$, where $I$ is the moment of inertia about the $z$-axis of rotation. Its spectrum is also one of SUSY as shown in Fig.\ 4, pairs of non-zero $m$ states being degenerate, with only the lowest $m=0$ state non-degenerate. The wave functions, to within a numerical factor, are $\exp (im\phi)$. Again, although a familiar example, the simple SUSY construction below has not received attention. (The SUSY of the rotor has been noted in different contexts \cite{ref10}, including in re-casting the one-dimensional hydrogen atom as a rotor \cite{ref14}.) We can construct the operators

\begin{equation}
Q=L_z \mathcal{T}/\sqrt{2I}, Q^{\dagger} = \mathcal{T} L_z /\sqrt{2I} =-Q,
\label{eqn7}
\end{equation}
where $\mathcal{T}$ is the time-reversal operator. Again, Eq.~(\ref{eqn1}) is satisfied as well as all but the criterion 5 of SUSY. There is a pleasing symmetry between the two problems of the free particle and free rotor. Besides the obvious roles of linear momentum and angular momentum in the two, note the roles of parity and time reversal in $Q$. Also note again the product form of two operators that both commute with the Hamiltonian in constructing $Q$. The alternative of an operator $q$, chosen as a linear combination of $L_z$ and $L_z \mathcal{T}$ such that criterion 5 is also satisfied, is immediate. 
 
\begin{figure}
 \vspace{-1in}
 \hspace{-.65in}
\scalebox{1.8}{\includegraphics[width=2in]{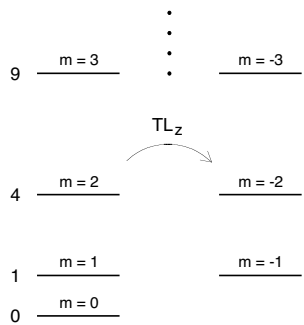}}
 \vspace{-1.3in}
\caption{Spectrum of a free rotor, displaying supersymmetry and the operator product of angular momentum and time-reversal.}
\end{figure}
 
This extended view of supersymmetry, and the employment of space-time symmetries in its construction, may find more applications and provide more insight into quantum mechanical problems.

I thank Richard Haymaker, Francois Gieres, Mikhail Plyushchay and C. V. Sukumar for discussions. This work has been supported by the National Science Foundation.


\begin{thebibliography}{}
\bibitem{ref9} The equivalence of these two points of view is not surprising. It reflects the alternatives of using the Cartesian Pauli spinors with squares equal to unity or the step-up/down combinations $\sigma^{\pm}$ whose squares vanish.
\end{thebibliography}
\end{document}